\documentclass{new_tlp}

\usepackage{todonotes}
\newcommand{\ynote}[1]{\todo[inline, color=green!40]{#1}}
\newcommand{\knote}[1]{\todo[inline, color=blue!10]{#1}}

\newtheorem{example}{Example}[section]
\newtheorem{definition}{Definition}[section]
\newtheorem{proposition}{Proposition}[section]
\newtheorem{corollary}{Corollary}[section]
\newtheorem{theorem}{Theorem}[section]
\newtheorem{lemma}{Lemma}[section]

\usepackage{mathptmx}
\usepackage[all]{xy}
\usepackage{comment}
\usepackage{amssymb}
\usepackage{amsfonts,amsmath,latexsym,amssymb}
\usepackage{tikz}
\usetikzlibrary{circuits.ee.IEC}
\usepackage{verbatim}
\usepackage{listings}
\usepackage{color}
 \usepackage{algpseudocode}
\usepackage{algorithm, url}
\usepackage{ enumerate }

\definecolor{LightGray}{gray}{0.8}

\newcommand{\Nat}{{\mathbb N}}

\newcommand{\cat}[1]{\ensuremath{\mathbf{#1}}}

\title[Productive Corecursion in Logic Programming]
{Productive Corecursion in Logic Programming\thanks{This work has been partially supported by EPSRC grant EP/K031864/1-2}}

\author[E. Komendantskaya and Y.Li]
{EKATERINA KOMENDANTSKAYA\\
Heriot-Watt University, Edinburgh, Scotland, UK\\
\email{ek19@hw.ac.uk}
\and
YUE LI\\
Heriot-Watt University, Edinburgh, Scotland, UK\\
\email{yl55@hw.ac.uk}
}

\begin{document}
\maketitle
\begin{abstract}
Logic Programming  is a Turing complete language. As a consequence, designing algorithms that decide termination and non-termination of programs or decide inductive/coinductive soundness of formulae is a challenging task. For example, the existing state-of-the-art algorithms can only semi-decide coinductive soundness of queries in logic programming for regular formulae. Another, less famous, but equally fundamental and important undecidable property is productivity.   If a derivation is infinite and coinductively sound, we may ask whether the computed answer it determines actually computes an infinite formula. If it does, the infinite computation is productive.  This intuition was first expressed under the name of computations at infinity in the 80s. In modern days of the Internet and stream processing, its importance lies in connection to infinite data structure processing. 

Recently, an algorithm was presented that semi-decides a weaker property -- of productivity of logic programs. A logic program is productive if it can give rise to productive derivations. In this paper we strengthen these recent results. We propose a method that semi-decides productivity of individual derivations for regular formulae. Thus we at last give an algorithmic counterpart to the notion of productivity of derivations in logic programming. This is the first algorithmic solution to the problem since it was raised more than 30 years ago. We also present an implementation of this algorithm. 
\end{abstract}

\section{Motivation}\label{sec:intro}

The traditional (inductive) approach
to Logic Programming (LP) is based on least fixed point semantics of
logic programs, and defines, for every logic program $P$, the
\emph{least Herbrand model} for $P$, i.e., the  set of
all (finite) ground terms {\em inductively entailed} by $P$.

\begin{example}[Natural numbers]\label{ex:nat} 
The program below defines the set of natural numbers:

\noindent
0. $\mathtt{nat(0)} \; \gets \;$\\ 
1. $\mathtt{nat(s(X))} \; \gets \; \mathtt{nat(X)}$


\noindent 
The least Herbrand model  comprises the terms
$\mathtt{nat(0), \ nat(s(0)),}$ $\mathtt{nat(s(s(0))), \ldots}$
\end{example}

The clauses of the above program can be viewed as inference rules
$\frac{}{\mathtt{nat(0)}}$ and
$\frac{\mathtt{nat(X)}}{\mathtt{nat(s(X))}}$, and the least Herbrand
model can be seen as the set obtained by the forward closure of these
rules. 
Some approaches to LP  are based
on this inductive view~\cite{HeintzeJ92} of programs.

In addition to viewing logic programs inductively, we can also view
them coinductively. The \emph{greatest complete Herbrand model} for a
program $P$ takes the backward closure of the rules derived from $P$'s
clauses, thereby producing the largest set of finite and infinite
ground terms \emph{coinductively entailed} by $P$. For example, the
greatest complete Herbrand model for the above program is the set containing all
of the finite terms in its least Herbrand model, together
with the term $\mathtt{nat(s(s(...)))}$ representing the first limit
ordinal. 

As it turns out, some logic programs have no natural inductive
semantics and should instead be interpreted coinductively:

\begin{example}[Streams of natural numbers]
\label{ex:natstream} 
The  next program  comprises the
clauses that define the natural numbers  and the following additional one that defines streams of natural numbers:

\vspace*{0.05in}

\noindent 
2. $\mathtt{nats(scons(X,Y))} \; \gets \; \mathtt{nat(X)},\mathtt{nats(Y)}$ 

\vspace*{0.05in}

\noindent 
No terms defined by $\mathtt{nats}$ are contained in the least
Herbrand model for this program, but its greatest complete Herbrand model
contains infinite terms representing infinite streams of natural numbers, like e.g. the infinite term $t = \mathtt{nats(scons(0,scons(0,
  \ldots)}$. 
\end{example}

Coinductive programs operating on infinite data structures are useful for reasoning about concurrent and non-terminating processes. For example, the program below shows a part of a description of concurrent behaviour of Internet servers~\cite{Dav01}:

\vspace*{0.05in}
\noindent 0. $\mathtt{resource([get(X)| In], [X|L]) \gets resource(In,L)}$\\
1. $\mathtt{resource([get(X)| In], []) \gets signal(novalue(get(X)))}$
\vspace*{0.05in}

\noindent The two clauses above describe how a server receives and processes the input streams of data. The usual list syntax $\mathtt{[\_ |\_ ]}$ of Prolog is used to denote the binary stream constructor $\mathtt{scons}$. The first clause 
describes the normal behaviour of the server that reads the input data, and the second allows to raise an exception (by signalling that no value was received).

SLD-resolution~\cite{Llo88} is an algorithm that allows to semi-decide whether a given formula is in the 
program's least Herbrand model.
In practice, SLD-resolution 
requires a logic program's derivations to be terminating in order for
them to be inductively sound. 
We might dually expect a logic program's non-terminating derivations
to compute terms in its greatest complete Herbrand model.  However,
non-termination does not play a role for coinduction dual to that
played by termination for induction. In particular, the fact that a
logic program admits non-terminating SLD-derivations does not, on its
own, guarantee that the program's computations completely capture its
greatest complete Herbrand model:

\begin{example}[Non-productive program]\label{ex:bad}
The following ``bad'' program gives rise to an infinite SLD-derivation:

\noindent
0. $\mathtt{bad(f(X))} \; \gets \; \mathtt{bad(f(X))}$

\noindent 
Although this program does not compute any infinite terms, the
infinite term $\mathtt{bad(f(f(...)))}$ is in its greatest complete
Herbrand model.
\end{example}

The problem here actually lies in the fact that  the  ``bad'' program fails to satisfy the
important property of productivity. The productivity requirement on
corecursive programs should reflect the
fact that an infinite computation can only be consistent with its
intended coinductive semantics if it is {\em globally productive},
i.e., if it actually produces an infinite object in the limit. 
This intuition lies behind the concept of \emph{computations at infinity} introduced in the
1980s~\cite{Llo88,EmdenA85}. The operational semantics of a
potentially non-terminating logic program $P$ was then taken to be the
set of all infinite ground terms computable by $P$ at infinity. For
example, the infinite ground term $t$ in Example~\ref{ex:natstream} is
computable at infinity starting with the query $
\mathtt{nats(X)}$. In modern terms, we would say that computations at infinity are \emph{(globally) productive computations}.


However, 
the notion of computations at infinity
does not by itself give rise to algorithms for semi-deciding coinductive entailment. 
Thirty years after the initial investigations into coinductive
computations, coinductive logic programming, implemented as CoLP, was
introduced \cite{GuptaBMSM07,SimonBMG07}.  CoLP provides practical
methods for terminating infinite SLD-derivations. 
CoLP's coinductive proof search is based on a
loop detection mechanism and unification without occurs check.
CoLP observes finite
fragments of SLD-derivations, checks them for unifying subgoals, and
terminates when loops determined by such subgoals are found. 

\begin{example}[Productive computation by SLD resolution]\label{ex:natstream2} 
The query
$\mathtt{nats(X)}$ to the program of 
Example~\ref{ex:natstream} gives rise to an SLD-derivation with a
sequence of subgoals $\mathtt{nats(X)} \leadsto^{X \mapsto scons(0,Y')} \mathtt{nats(Y')} \leadsto \ldots$.
Observing that
$\mathtt{nats(scons(0,Y'))}$ and $\mathtt{nats(Y')}$ unify (note the absence of occurs check)
and thus comprise a loop, CoLP concludes that $\mathtt{nats(X)}$ has
been proved 
and  returns the
answer $\mathtt{X=scons(0,X)}$ in the form of a ``circular'' term
indicating that this program logically entails the term $t$ in
Example~\ref{ex:natstream}.
\end{example}

CoLP is sound, but incomplete, relative to greatest complete Herbrand
models \cite{GuptaBMSM07,SimonBMG07}. But, perhaps surprisingly, it is
{\em neither} sound {\em nor} complete relative to computations at
infinity. CoLP is not sound because our ``bad'' program from
Example~\ref{ex:bad} computes no infinite terms at infinity for the
query $? \gets \mathtt{bad(X)}$, whereas CoLP notices a loop and
reports success. CoLP is not complete because not all terms computable at
infinity by all programs can be inferred by CoLP. In fact, CoLP's loop
detection mechanism can only terminate if the term computable at
infinity is a \emph{regular}
term~\cite{Courcelle83,JaffarS86}. Regular terms are terms that can
be represented as trees that have a finite number of distinct
subtrees, and can therefore be expressed in a closed finite form
computed by circular unification. The ``circular'' term $\mathtt{X}$ =
$\mathtt{scons(0,X)}$ in Example~\ref{ex:natstream2} is so expressed.
For irregular terms (e.g. expressing a stream of Fibonacci numbers, cf. Example~\ref{ex:fibs}), CoLP simply does not terminate. 

The upshot is that  the loop detection method of CoLP cannot faithfully capture the operational
meaning of computations at infinity.  In this paper, we propose a solution to this problem, by combining loop detection and productivity within one framework.

\section{Results of This Paper by Means of an Example}\label{sec:ex}

We return to our ``Server" example, but this time take only the clause that describes its normal execution (without exceptions):\\
$\mathtt{resource([get(X)| In], [X|L]) \gets resource(In,L)} \ \ \ \  \ \ \ \ (*)$

\noindent The second argument of $\mathtt{resource}$ is the input stream received by the server, and its first argument is the stream of successfully received and 
read data.
We can take e.g. the query\\  $\mathtt{resource(X,Y), zeros(Y)}$, asking the server to accept as input the stream of zeros defined as
$\mathtt{zeros([0|X]) \gets zeros(X)}$.

Assuming a fair selection of subgoals in a derivation, we will have the following SLD-derivation\\ 
$\mathtt{\underline{resource(X,Y)}, \underline{\underline{zeros(Y)} }}\leadsto^{X \mapsto [get(X')|In], Y \mapsto [X'|L]}\leadsto \mathtt{\underline{resource(In,L)}, zeros(X'|L)} \leadsto^{X'\mapsto 0} \leadsto$ \\ $ \mathtt{\underline{resource(In,L)}, \underline{\underline{zeros(L)}}} \leadsto \ldots $\\
and the substitution 
$X \mapsto \mathtt{[get(0)| get(0)| \ldots]}$ will be computed at infinity.
This regular computation will be processed successfully by the loop detection method of CoLP (relying on a unification algorithm without occurs check). We underlined the loops above.

There are three cases where CoLP fails to capture the notion of productive computations: 

 \emph{Case 1. The coinductive definition does not contain constructors of the infinite data structure.} Imagine  we have the clause $\mathtt{resource(In, L) \gets resource(In,L)}$ instead of (*). This new clause simply asserts a tautology:  a server receives the data when it receives the data.
Querying again $\mathtt{resource(X,Y), zeros(Y)}$, we will get an infinite looping SLD-derivation, that however will not compute an infinite ground term at infinity (the first argument will not be instantiated):\\
$\mathtt{\underline{resource(X,Y)}, \underline{\underline{zeros(Y)} }}\leadsto^{id}\leadsto \mathtt{\underline{resource(X,Y)}, zeros(Y)} \leadsto^{Y\mapsto [0|Y']} \leadsto$  \\ $ \mathtt{\underline{resource(X,[0|Y'])}, \underline{\underline{zeros(Y')}}} \leadsto \ldots $
\vspace*{0.05in}

\emph{Case 2. The coinductive definition contains fresh variables that do not allow to accumulate composition of substitutions in the course of a derivation.}
Imagine clause (*) is replaced by: $\mathtt{resource([get(X)| In], [X|L]) \gets resource(Z,L)}$ Then we would still have an infinite looping derivation, but it will not allow us to meaningfully compose the computed substitutions in the first argument:

\noindent $\mathtt{\underline{resource(X,Y)}, \underline{\underline{zeros(Y)} }}\leadsto^{X \mapsto [get(X')|In], Y \mapsto [X'|L]}\leadsto \mathtt{\underline{resource(Z,L)}, zeros(X'|L)} \leadsto^{X'\mapsto 0} \leadsto$ \\ $ \mathtt{\underline{resource(Z,L)}, \underline{\underline{zeros(L)}}} \leadsto^{Z \mapsto [get(X'')|In'], L \mapsto [X''|L']}\leadsto \ldots  $

\noindent Note that in the last step the computed substitution for the fresh variable $Z$ does not affect the substitution $X \mapsto [get(X')|In]$. However, the loops will still be detected as shown.

\vspace*{0.05in}
\emph{Case 3. The infinite data structure is defined via a circular unification, rather than computed by an infinite number of derivation steps.}
Imagine that we force our definition to always produce an infinite stream in its first argument by defining:\\
$\mathtt{resource([get(X)| In], In, [X|L]) \gets resource(In,In,L)}$.

\noindent We can have the following derivation (no need to give a stream of zeros as an input):

\vspace*{0.05in}
\noindent $\mathtt{\underline{resource(X,Y,[0|Z])}}\leadsto^{X \mapsto [get(X')|In], Y\mapsto In, Z \mapsto L, X' \mapsto 0}
\leadsto \mathtt{\underline{resource(In,In,L)}} \leadsto^{In'\mapsto [get(X'')|In'] \ldots \infty} $
\vspace*{0.05in}

\noindent Here, we will not have an infinite SLD-derivation, as the last step fails the occurs check.
But CoLP's loop detection without occurs check will terminate successfully, as the underlined loop is found and the looping terms will unify by circular unification (denoted~by~$\infty$).



In all of the above three cases, if the three programs share the common signature of (*), their greatest complete Herbrand models will  contain the term 
$\mathtt{resource([get(0)| get(0)| \ldots],[0|0|\ldots])}$.  In fact, the three looping derivations allude to this term when they succeed by the loop detection without occurs check. However, as we have seen from these examples, in neither of the three derivations the loop detection actually guarantees that there is a way to continue the SLD-derivation lazily in order to compute this infinite ground term at infinity.
Thus, in all three cases, the loop detection method is unsound relative to computations at infinity.

\emph{The question we ask is: Assuming we can guarantee that none of the three cases will occur in our derivations, can the
 loop detection method serve as an algorithm for semi-deciding whether a derivation is productive, or equivalently, whether an infinite term is computable at infinity?}

In this paper, we answer this question in the positive. 
Case 2 can be eliminated by a simple syntactic check disallowing fresh (or ``existential'') variables in the bodies of the clauses. We call the resulting programs \emph{universal}.
Case 1 is more subtle. In logic programming setting, unlike for example functional languages, it is not easy to identify which of the clauses form which inductive definitions by which constructors.
Such properties are usually not  decided until  run time. 
Consider the following example. 

\begin{example}[Difficulty in detection of constructor productivity in LP]\label{ex:nonprod}
The following program \\
0. $\mathtt{p(s(X1),X2,Y1,Y2)} \, \gets \,  \mathtt{q(X2,X2,Y1,Y2)}$ \\
1. $\mathtt{q(X1,X2,s(Y1),Y2)} \, \gets \, \mathtt{p(X1,X2,Y2,Y2)}$\\
seemingly defines, by mutual recursion, two coinductive predicates $\mathtt{p}$ and $\mathtt{q}$, with constructor $\mathtt{s}$. However, the SLD-derivation for a query $\mathtt{p(s(X),s(Y),s(Z),s(W))}$  will not produce an infinite term at infinity. However, it will produce loops that will be found by CoLP!
\end{example}
 
As a solution, we propose to use the notion of \emph{observational productivity} suggested recently by~\cite{KJS17,FK16}.
Given a logic program $P$, a query $A$,  and an SLD-derivation for $P$ and $A$, we can analyse the structure of this derivation and detect which of the unifiers used in its course are most general unifiers (mgus) and which are most general matchers (mgms). Systematic analysis of such steps is called \emph{structural resolution} in~\cite{JKK15}.
We define a logic program to be \emph{observationally productive} if it is impossible to construct an infinite derivation only by mgms for it. For ``good" coinductive programs like e.g. the one of Example~\ref{ex:natstream}, infinite SLD-derivations exist, but not infinite derivations by mgms: the derivations by mgms terminate as soon as they run out of coinductive constructors to match against, and then derivations by mgus produce further substitutions and thus produce more constructors. Examples~\ref{ex:nat-red}, ~\ref{ex:nonp} and ~\ref{ex:sres} will make this intuition clear.  In~\cite{KJS17}, an algorithm for checking observational productivity of logic programs was introduced. In this paper, we take this  algorithm  as a sufficient formal 
check for ruling out programs giving rise to coinduction without constructors, as described in Case 1 above. 

Addressing Case 3 requires modifications to the loop detection algorithm: it should be able to form circular substitutions for cases like (*), while ruling out cases like Case 3. 

This paper thus establishes two results for universal and observationally productive programs:

		\emph{(1) It shows that non-termination of derivations for such programs guarantees computation of an infinite term at infinity.} 
	In terms of our ``Server" example and its clause (*), if computations continue indefinitely, we know that the server receives and processes an infinite stream of data;
	
	\emph{(2) It proposes a novel loop detection algorithm that guarantees production of an infinite term at infinity.} 
	In the ``Server" example terms, if the loop detection algorithm succeeds, we know how  the productive computation can proceed  infinitely long.

The novelty of these results is three-fold:\\
\emph{-- theoretically}, it is the first time that non-terminating observationally productive derivations are proven to be globally productive (i.e. sound relative to computations at infinity);\\
\emph{-- practically}, it presents the first algorithm for semi-deciding computation of infinite terms at infinity since the notion was introduced in the 80s;\\
\emph{-- methodologically}, all proofs employ the methods of structural resolution, which  extends the existing methodological machinery of LP and allows to achieve results that are not directly provable for the SLD-resolution.


The paper proceeds as follows.  Section~\ref{sec:backgr} gives all background definitions alongside a modified version of structural resolution that has not appeared in the literature before.
This section also proves that this variant of structural resolution is sound and complete relative to SLD-resolution. This allows us to switch freely between SLD- and structural resolution throughout the paper. 
  Section~\ref{sec:models} proves soundness and completeness of infinite structural resolution derivations relative to computations at infinity, for universal and observationally productive programs. As a corollary, it gives conditions for soundness of infinite SLD-derivations relative to computations at infinity. 
	Section~\ref{sec:co-S} introduces the novel loop detection algorithm for structural resolution, and proves its soundness relative to computations at infinity. 
In Section~\ref{sec:conclusion} we  conclude this paper.
Implementation of the  algorithm of Section~\ref{sec:co-S} is available at\\ \url{https://github.com/coalp/Productive-Corecursion} and is discussed in Appendix A.6. 
\section{Background: S-resolution and Observational Productivity}\label{sec:backgr}

In this section we 
introduce structural resolution by means of an operational (small-step) semantics. To enable
the analysis of infinite terms, we adopt the
standard definitions of first-order terms as trees~\cite{Courcelle83,JaffarS86,Llo88}. But, unlike earlier 
approaches~\cite{JKK15}, we avoid analysis of (SLD-)derivation trees in this paper 
and work directly with S-resolution reductions. 

We write $\Nat^*$ for the set of all finite words over the set $\Nat$
of natural numbers. The length of $w\in\Nat^*$ is denoted $|w|$. The
empty word $\epsilon$ has length $0$; we identify $i \in \Nat$ and the
word $i$ of length $1$.  
A set $L \subseteq \Nat^*$ is a \emph{(finitely branching)
 tree
  language} provided: i) for all $w \in \Nat^*$ and all $i,j \in
\Nat$, if $wj \in L$ then $w \in L$ and, for all $i<j$, $wi \in L$;
and ii) for all $w \in L$, the set of all $i\in \Nat$ such that $wi\in
L$ is finite.  A non-empty tree language always contains $\epsilon$,
which we call its {\em root}.  
 A tree language is {\em finite} if it is a finite subset of $\Nat^*$, and {\em infinite}
otherwise. 

A \emph{signature} $\Sigma$ is a non-empty set of \emph{function
  symbols}, each with an associated arity. 
To define terms over $\Sigma$, we
assume a countably infinite set $\mathit{Var}$ of {\em variables}
disjoint from $\Sigma$, each with arity $0$. 
If $L$ is a
non-empty tree language and $\Sigma$ is a signature, then a
\emph{term} over $\Sigma$ is a function $t: L \rightarrow \Sigma \cup
\mathit{Var}$ such that, for all $w\in L$, $\mathit{arity}(t(w)) =
\;\,\mid \!\{i \mid wi \in L\}\!\mid$.  Terms are finite or infinite
if their domains are finite or infinite.  A term $t$ has a
depth $\mathit{depth}(t) = 1+ \max\{|w| \mid \, w\in L\}$.  



\begin{example}[Term tree]\label{ex:term-tree}
Given $L = \{ \epsilon, 0, 00, 01 \}$, the atom
$\mathtt{stream(scons(0,Y))}$ can be seen as the term tree $t$ given
by the map $t(\epsilon) = \mathtt{stream}$, $t(0) = \mathtt{scons}$,
$t(00) = \mathtt{0}$, $t(01) = \mathtt{Y}$.
\end{example}

The set of finite (infinite) terms over a
signature $\Sigma$ is denoted by $\mathbf{Term}(\Sigma)$
($\mathbf{Term}^\infty(\Sigma)$). The set of {\em all} (i.e., finite
{\em and} infinite) terms over $\Sigma$ is denoted by
$\mathbf{Term}^\omega(\Sigma)$. Terms with no occurrences of variables
are \emph{ground}. We write $\textbf{GTerm}(\Sigma)$
($\textbf{GTerm}^\infty(\Sigma)$, $\mathbf{GTerm}^\omega(\Sigma)$) for
the set of finite (infinite, {\em all}) ground terms over $\Sigma$.

A \emph{substitution} over $\Sigma$ is a total function $\sigma:
\mathit{Var} \to \mathbf{Term}^{\omega}(\Sigma)$. 
Substitutions are extended from variables
to terms homomorphically. 
We write $\mathit{id}$ for the identity substitution. 
Composition of substitutions is denoted by juxtaposition.  Composition
is associative, so we write $\sigma_3\sigma_2\sigma_1$ rather than
$(\sigma_3\sigma_2)\sigma_1$ or $\sigma_3(\sigma_2\sigma_1)$.

A substitution $\sigma$ is a \emph{unifier}
for $t, u \in \cat{Term}(\Sigma)$ if $\sigma(t) = \sigma(u)$, and is a
\emph{matcher} for $t$ against $u$ if $\sigma(t) = u$.  If $t, u \in
\cat{Term}^\omega(\Sigma)$, then we say that $u$ is an {\em instance}
of $t$ if $\sigma(t) = u$ for some $\sigma$.
A substitution $\sigma_1$ is {\em more general} than a substitution
$\sigma_2 $
if there exists a substitution $\sigma$
such that $\sigma \sigma_1(\mathtt{X}) = \sigma_2(\mathtt{X})$ for
every $\mathtt{X} \in \mathit{Var}$. A substitution $\sigma$ 
is a {\em most general unifier} ({\em mgu}) for
$t$ and $u$, denoted $t \sim_\sigma u$, if it is a unifier for $t$ and
$u$ and is more general than any other such unifier. A {\em most
  general matcher} ({\em mgm}) $\sigma$ for $t$ against $u$, denoted
$t \prec_\sigma u$, is defined analogously. Both mgus and mgms are
unique up to variable renaming if they exist.  
In many unification algorithms, the \emph{occurs check} condition is
imposed, so that mappings $\mathtt{X} \mapsto
t[\mathtt{X}]$, where $t[\mathtt{X}]$ is a term containing
$\mathtt{X}$, are disallowed. 
We will assume that mgus and mgms are computed by any standard unification algorithm~\cite{Llo88} with occurs check, unless otherwise stated.

A clause $C$ over $\Sigma$  
is given by $A \gets B_0, \ldots , B_n$ where the head $A \in \mathbf{Term}(\Sigma)$ and the \emph{body} $B_0, \ldots  B_n$ is a list of terms in $\mathbf{Term}(\Sigma)$. 
Throughout the paper, we refer to standard definitions of the least and greatest complete Herbrand models
and recall them in Appendix A.1.
  



Following~\cite{FK16}, we 
distinguish several  kinds of
reductions for LP: 

\begin{definition}[Different kinds of reduction in LP]
If $P$ is a logic program and $A_1, \ldots, A_n$ are atoms, then
\begin{itemize}
\item {\em SLD-resolution reduction}:\\ $[A_1, \ldots, A_i, \ldots ,
  A_n] \leadsto_P [\sigma(A_1), \ldots, \sigma(A_{i-1}), \sigma(B_0),
  \ldots, \sigma(B_m), \sigma(A_{i+1}), \ldots, \sigma(A_n)]$\\ if $A
  \gets B_0 , \ldots , B_m \in P$ and $A_i \sim_{\sigma} A$.


\item {\em rewriting reduction}:\\ $[A_1, \ldots , A_i , \ldots ,
  A_n] \rightarrow_P [A_1, \ldots, A_{i-1},\sigma(B_0), \ldots,
  \sigma(B_m), A_{i+1}, \ldots, A_n]$ \\ if $A \gets B_0 , \ldots , B_m
  \in P$ and $A \prec_{\sigma} A_i$.


\item {\em substitution reduction}:\\ $[A_1, \ldots , A_i , \ldots,
  A_n] \hookrightarrow_P [\sigma(A_1), \ldots, \sigma(A_i), \ldots ,
  \sigma(A_n)]$ \\ if $A \gets B_0 , \ldots , B_m \in P$ and $A
  \sim_{\sigma} A_i$ but not $A \prec_{\sigma} A_i$.
	
	In each of the above three cases, we will say that $A_i$ \emph{resolves against the clause} $A \gets B_0 , \ldots , B_m
  \in P$.

	

\end{itemize}

We may omit explicit mention of $P$
as 
a subscript on reductions when it is clear from
context. We write $\rightarrow^n$ to denote rewriting by {\em at
  most} $n$ steps of $\rightarrow$, where $n$ is a natural number.
	We use similar notations for $\leadsto$ and $\hookrightarrow$ as
required.
We assume, as is standard in LP, that all variables are \emph{standardised apart} when terms are matched or unified against the program clauses.
\end{definition}

	
If $r$ is any reduction relation, 
we will call
any (finite or infinite) sequence of $r$-reduction steps an {\em$r$-derivation}. 
An $r$-derivation is called an \emph{$r$-refutation} if its last goal is the empty list. 
An SLD-resolution derivation is {\em fair} if either it is finite, or
it is infinite and, for every atom $B$ appearing in some goal in the
SLD-derivation, (a further instantiated version of) $B$ is chosen
within a finite number of steps.

\begin{example}[SLD- and rewriting reductions]\label{ex:nat-red}

The following are SLD-resolution and rewriting
derivations, respectively, with respect to the program of Example~\ref{ex:natstream}:
\begin{itemize}
\item $[\mathtt{nats(X)}] \leadsto [\mathtt{nat(X')},
  \mathtt{nats(Y)}] \leadsto [\mathtt{nats(Y)}] \leadsto
  [\mathtt{nat(X'')}, \mathtt{nats(Y')}]\leadsto \ldots$

\item $[\mathtt{nats(X)}] $
\end{itemize}
The reduction relation $\leadsto_P$ models
traditional SLD-resolution steps \cite{Llo88} with respect to $P$.
Note that for this program, SLD-resolution derivations are infinite, but rewriting derivations are always finite.
\end{example}

The observation that, for some coinductive programs,
$\rightarrow$ reductions are finite and thus can serve as
measures of finite observation, has led to the following definition of
observational productivity in LP, first introduced in~\cite{KPS12-2}:

\begin{definition}[Observational productivity]\label{def:prod}
A program $P$ is {\em observationally productive}
if every rewriting
derivation with respect to $P$ is finite.
\end{definition}
\vspace*{-0.1in}

\begin{example}[Observational productivity]\label{ex:nonp}
The program from Example~\ref{ex:natstream}  is observationally
productive, whereas the program from Example~\ref{ex:bad} -- not,
as we have a rewriting derivation:
$\mathtt{bad(f(X))} \rightarrow \mathtt{bad(f(X))}\rightarrow \mathtt{bad(f(X))} \rightarrow \ldots  $. 
\end{example}

Because rewriting derivations are incomplete, they can be combined with substitution reductions to achieve completeness,
which is the main idea behind the \emph{structural resolution} explored in~\cite{KPS12-2,JKK15,FK16}.
Below, we present yet another version of combining the two kinds of reductions:

\begin{definition}[S-resolution reduction]
Given a productive program $P$, we define {\em S-resolution   reduction}: \\
		$ [A_1, \ldots ,   A_n] 
\rightarrow^{n}_P [B_1, \ldots, B_i, \ldots , B_m] \hookrightarrow_P [\theta(B_1), \ldots, \theta(B_i) , \ldots,  \theta(B_m)] \rightarrow_P $\\
$[\theta(B_1), \ldots, \theta(B_{i-1}),\theta(C_1), \ldots, \theta(C_k), \theta(B_{i+1}),\ldots,  \theta(B_m)]$,\\ where $C \gets C_1, \ldots , C_k$ is a clause in $P$, and
$C \sim_{\theta} B_i$. We will denote this reduction by\\  $[A_1, \ldots,
 A_n] \leadsto^S_P [\theta(B_1), \ldots, \theta(B_{i-1}),\theta(C_1), \ldots, \theta(C_k), \theta(B_{i+1}),\ldots,  \theta(B_m)]$.
\end{definition}

\begin{example}[S-resolution reduction]\label{ex:sres}
Continuing Example~\ref{ex:nat-red}, S-resolution derivation for that logic program is given by:\\
 $[\mathtt{nats(X)}]
  \hookrightarrow [\mathtt{nats(scons(X',Y))}] \rightarrow
                 [\mathtt{nat(X')}, \mathtt{nats(Y)}]
                 \hookrightarrow [\mathtt{nat(0)}, \mathtt{nats(Y)}]$\\
                 $\rightarrow [\mathtt{nats(Y)}]
                 \hookrightarrow [\mathtt{nats(scons(X'',Y'))}]
                 \rightarrow \ldots$

\noindent
The
initial sequences of the SLD-resolution (see Example~\ref{ex:nat-red}) and S-resolution reductions shown above
each compute the partial answer $\{\mathtt{X} \mapsto
\mathtt{scons(0,scons(X'', Y'))}\}$ to the query $\mathtt{nats(X)}$.
\end{example}

Several formulations of S-resolution exist in the literature \cite{KPS12-2,JKK15,FK16}, some of them are incomplete relative to SLD-resolution.
However, the above definition is complete:

\begin{theorem}[Operational equivalence of terminating S-resolution and SLD-resolution]\label{th:equiv}
Given a logic program $P$ and a goal $A$, there is an SLD-refutation for $P$ and $A$ iff there is an S-refutation for $P$ and $A$. 
\end{theorem} \vspace*{-0.1in}
\begin{proof}
Both parts of the proof proceed by induction on the length of the SLD- and S-refutations.
The proofs rely on one-to-one correspondence between SLD- and S-derivations.  Suppose $D$ is an SLD-derivation, then we can construct a corresponding S-derivation  $D^*$ as follows. 
 Some reductions in $D$ in fact compute mgms,  these reductions are modelled by rewriting reductions in $D^*$ directly. Those reductions that involve proper mgus in $D$ are modelled by
composition of two reduction steps $\rightarrow \circ \hookrightarrow$  in $D^*$.
\end{proof}
A corollary of this theorem is inductive soundness and completeness of S-refutations relative to the least Herbrand models.

\section{Soundness of Infinite S-Resolution Relative to SLD-Computations at Infinity}\label{sec:models} 

A first attempt to give an operational semantics corresponding to
greatest complete Herbrand models of logic programs was captured by
the notion of \emph{computations at infinity} for SLD-resolution
\cite{EmdenA85,Llo88}. 
Computations at infinity are usually given
relative to an ultrametric on terms, constructed as follows.

We define the {\em truncation} of a term $t \in \mathbf{Term}^\omega(\Sigma)$ at depth $n \in \Nat$, denoted by $\gamma\: '(n,t)$. We introduce a new
nullary symbol $\diamond$ to denote the leaves of truncated branches.

\begin{definition}[Truncation of a term]\label{df:trunc}
A \emph{truncation}  is a mapping $\gamma\:':
\Nat \times \mathbf{Term}^\omega(\Sigma) \rightarrow
\mathbf{Term}(\Sigma \cup \diamond)$ where, for every $n \in \Nat$ and $t \in \mathbf{Term}^\omega(\Sigma)$, the term 
$\gamma\:'(n,t)$ is constructed as follows:\\ 
(a) the domain $\mathit{dom}(\gamma\:'(n,t))$ of the term $\gamma\:'(n,t)$ is $\{m \in \mathit{dom}(t) \mid  |m|\leq n \}$; \\
(b) 
\[ \gamma\:'(n,t)\ (m) =
  \begin{cases}
    t(m)       & \quad \text{if } |m| < n\\
    \diamond   & \quad \text{if } |m| = n\\
  \end{cases}
\]

\noindent	
For $t,s \in \mathbf{Term}^\omega(\Sigma)$, we define $\gamma(s,t) =
min\{n \;| \ \gamma\:'(n,s) \neq \gamma\:'(n,t)\}$, so that $\gamma(s,t)$
is the least depth at which $t$ and $s$ differ.  If we further define
$d(s,t) = 0$ if $s = t$ and $d(s,t) = 2^{-\gamma(s,t)}$ otherwise,
then $(\mathbf{Term}^\omega(\Sigma), d)$ is an ultrametric space.
\end{definition}

The definition of SLD-computable at infinity relative to a given
ultrametric was first given in~\cite{Llo88}, we extend it here and redefine this
 with respect to an arbitrary infinite term, rather than a ground infinite term:

\begin{definition}[Formulae SLD-computable at infinity]\label{def:SLD-comp-inf}
 The term tree $t^\infty \in
\cat{Term}^\infty(\Sigma)$ is \emph{SLD-computable at infinity} with
respect to a program $P$ if there exists a term tree $t \in
\cat{Term}(\Sigma)$ and an infinite fair SLD-resolution derivation $G_0
= t \leadsto G_1 \leadsto G_2 \leadsto \ldots G_k \leadsto \ldots$ with mgus $\theta_1, \theta_2,
\ldots, \theta_k, \ldots$ such that $d(t^\infty, \theta_k\ldots \theta_1(t))
\rightarrow 0$ as $k \rightarrow \infty$. If such a $t$ exists, we
say that $t^\infty$ is SLD-computable at infinity with respect to $t$. 
\end{definition}
Note the fairness requirement above. Defining $C_P = \{ t^\infty \in
\mathbf{GTerm}^\infty(\Sigma)\, |\, t^\infty$ is SLD-computable at infinity
with respect to a program $P$ by some $t \in \cat{Term}(\Sigma)\}$, we have
that
$C_P$ is a subset of the greatest complete Herbrand model of $P$~\cite{EmdenA85,Llo88}.






\begin{example}[Existential variables and SLD computations at infinity]\label{ex:py}
Consider the following program that extends Example~\ref{ex:natstream}.  

\noindent
3. $\mathtt{p(Y) \gets nats(X)}$

\noindent Although an infinite term is SLD-computable at infinity with respect to $\mathtt{nats(X)}$,
no infinite instance of
$\mathtt{p(Y)}$ is SLD-computable at infinity. Nevertheless, $\mathtt{p(0)}$ and other instances of
$\mathtt{p(Y)}$ are logically entailed by this program and are in its greatest complete Herbrand model. 

Our Case 2 of Section~\ref{sec:ex} is also an example of the same problem.
\end{example}

To avoid such problems, we introduce a restriction on the shape of clauses and work only with logic programs
in which only variables occurring in the heads of clauses can occur in their bodies.
Formally, for each program clause $C \gets C_1, \ldots , C_n$, we form the set $FV(C)$ of all free variables in $C$, and similarly the set $FV(C_1, \ldots , C_n)$ of all free variables in $C_1, \ldots , C_n$. We require that $FV(C_1, \ldots , C_n) \subseteq FV(C)$ and call such logic programs \emph{universal logic programs}.

We now establish an important property -- that if a program is observationally productive and universal, then it necessarily gives rise to globally productive S-resolution derivations.
We call a substitution $\theta$ for a variable $X$ \emph{trivial} if it is a  renaming or identity mapping, otherwise it is \emph{non-trivial for $X$}. A substitution $\theta$ is non-trivial for a term $t$ with variables $X_1, \ldots, X_n$ if $\theta$ is non-trivial for at least one  variable $X_i \in \{X_1, \ldots, X_n\}$. We will call a composition $\theta_n  \ldots \theta_1$ non-trivial for $t$ if $\theta_1$ is non-trivial for $t$ and for each $1< k\leq n,$   $\theta_{k}$ is non-trivial for term $\theta_{k-1}
  \ldots \theta_1(t)$.

\begin{lemma}[Productivity Lemma]\label{lem:sres}
Let $P $ be an observationally productive and universal program 
and   let $t \in
\mathbf{Term}(\Sigma)$. 
Let $D$ be an infinite S-resolution derivation given by $G_0=t \leadsto^S G_1 \leadsto^S G_2 \leadsto^S \ldots$, then
 for every $G_i \in D$, there is a $G_j \in D$, with $j > i$, such that, given computed mgus $\theta_i, \ldots, \theta_1$ up to $G_i$ and the computed mgus $\theta_j, \ldots, \theta_1$ up to $G_j$, $d(t^\infty,\theta_i, \ldots, \theta_1(t)) > d(t^\infty,\theta_j, \ldots, \theta_1(t))$,
for some term $t^\infty \in \mathbf{Term}^{\infty}(\Sigma)$.
\end{lemma}
\begin{proof}[Proof Sketch]
The full proof is given in Appendix A.2, and proceeds by showing that universal and productive programs have to 
produce non-trivial substitutions in the course of S-resolution derivations. Non-trivial computed substitutions contribute to construction of the 
infinite term at the limit. The construction would have been a mere adaptation of the limit term construction of~\cite[p. 177]{Llo88}, had we not extended 
the notion of SLD-computable at infinity to all, not just ground, terms. This extension required us to redefine the limit term construction substantially, and use the properties of observationally productive S-resolution reductions.  
\end{proof}

Note how both requirements of universality and productivity are crucial in the above lemma.
For non-productive programs as in Example~\ref{ex:bad} or in Case 1 of Section~\ref{sec:ex}, an infinite sequence of rewriting steps will not produce any substitution. Analysis of Case 2 in Section~\ref{sec:ex} explains 
 the importance of the universality condition.


We now state the first major result: that for productive and universal logic programs, S-resolution derivations are sound and complete relative to SLD-computations at infinity.
First, we extend the notion of a fair derivation to S-resolution. An S-resolution derivation is {\em fair} if either it is finite, or
it is infinite and, for every atom $B$ appearing in some goal in the
S-resolution derivation, (a further instantiated version of) $B$ is resolved against some program clause 
within a finite number of steps.

\begin{theorem}[Soundness and completeness of observationally productive S-resolution]\label{th:ss}
Let $P $ be an observationally productive and universal program, and 
  let $t \in
\mathbf{Term}(\Sigma)$.

There is an infinite fair S-resolution derivation for $t$ iff there is a $t' \in
\cat{Term}^\infty(\Sigma)$, such that $t'$ is SLD-computable at infinity with respect to $t$.


\end{theorem}
\begin{proof}[Proof Sketch]
The full proof is in Appendix A.3. It first proceeds by coinduction to show a one-to-one correspondence between an infinite SLD-derivation and an infinite S-derivation. The argument is very similar to the proof of Theorem~\ref{th:equiv}. The rest of the proof uses Productivity Lemma~\ref{lem:sres} to show that an infinite derivation must produce an infinite term as a result.
\end{proof}

The practical significance of the above theorem  is in setting the necessary and sufficient conditions for guaranteeing that, given an infinite fair S-resolution reduction, we are guaranteed that it will compute an infinite term at infinity. We emphasise this consequence in a corollary:

\begin{corollary}[Global productivity of infinite S-resolution derivations]
Let $P $ be an observationally productive and universal program, and 
  let $t \in
\mathbf{Term}(\Sigma)$.

If there is an infinite fair S-resolution derivation $G_0
= t \leadsto^S G_1 \leadsto^S G_2 \leadsto^S \ldots G_k \leadsto^S \ldots$ with mgus $\theta_1, \theta_2,
\ldots \theta_k \ldots$ then there exists $t^\infty \in \cat{Term}^\infty(\Sigma)$ such that $d(t^\infty, \theta_k\ldots \theta_1(t))
\rightarrow 0$ as $k \rightarrow \infty$.
 
\end{corollary}

As a corollary of soundness of SLD-computations at infinity~\cite{Llo88} and the above result, we obtain that fair and infinite S-resolution derivations 
are sound relative to complete Herbrand models, given an observationally productive and universal program $P$.
Another important corollary that follows from the construction of the proof above guarantees that, if a program is observationally productive and universal, then any infinite fair SLD-resolution derivation for it will result in computation of an infinite term at infinity:
\begin{corollary}[Global productivity of infinite SLD-resolution derivations]
Let $P $ be an observationally productive and universal program, and 
  let $t \in
\mathbf{Term}(\Sigma)$.

If there is an infinite fair SLD-resolution derivation for $t$ then there is a $t^\infty \in
\cat{Term}^\infty(\Sigma)$, such that $t^\infty$ is SLD-computable at infinity with respect to $t$.
\end{corollary}



This section has established that infinite derivations for universal and observationally productive programs \emph{``cannot go wrong"}, in the sense 
that they are guaranteed to be globally productive and compute infinite terms. This is the first time a result of this kind is proven in LP literature. 
Although sound, infinite S- or SLD-resolution derivations do not provide an implementable  procedure for semi-deciding coinductive 
entailment.  We address this problem in the next section.
\section{Co-S-resolution}\label{sec:co-S}
In this section we embed a loop detection algorithm in S-resolution derivations, and thus obtain an algorithm of \emph{co-S-resolution} that
  can semi-decide whether an infinite term is SLD-computable at infinity for observationally productive and universal programs. 
To achieve this,  we refine the loop detection method of co-SLD resolution~\cite{GuptaBMSM07,SimonBMG07,AnconaCoSLD15} which we recall in Appendix A.4 for convenience.
We use  $\approx$ to denote unification without occurs check, 
see e.g.~\cite{CCT82} for the algorithm. 
We use 
the notational style of~\cite{AnconaCoSLD15} and  introduce a set $S_i$ for each predicate $A_i$ in a goal, where $S_i$ records the atoms 
from previous goals whose derivation depends on the derivation of $A_i$. We call $S_i$ the \emph{ancestors set} for $A_i$. 




\begin{definition}[Algorithm of Co-S-Resolution]\label{df:co-S}
\begin{itemize}
	\item  \emph{rewriting reduction ($G\rightarrow G'$):} Let $G=[(A_1,S_1),\ldots,(A_n, S_n)]$. If $B_0\prec_\theta A_k$ for some program clause $B_0\leftarrow B_1,\ldots,B_m$ and some $k$, then let $S'=S_k\cup \{A_k\}$. Then we derive \[G'=[ (A_1, S_1),\ldots,(A_{k-1}, S_{k-1}),(\theta(B_1), S'),\ldots,(\theta(B_m), S'),(A_{k+1}, S_{k+1}),\ldots,(A_n,S_n)]\] 
\item \emph{substitution reduction ($G\hookrightarrow G'$):} Let $G=[(A_1,S_1),\ldots,(A_n, S_n)]$. If $B_0\sim_\theta A_k$ but not \\$B_0\prec_\theta A_k$ for some program clause $B_0\leftarrow B_1,\ldots,B_m$ and some $k$. Then we derive \[G'=\theta \big(\ [ (A_1, S_1),\ldots,(A_n,S_n)]\ \big)\]
\item \emph{S-reduction ($G\leadsto^S G'$):}  $G\rightarrow^n [(A_1,S_1),\ldots,(A_n, S_n)]\hookrightarrow\theta \big(\ [ (A_1, S_1),\ldots,(A_n,S_n)]\ \big)\rightarrow $ \[G'=\theta \big(\ [ (A_1, S_1),\ldots,(A_{k-1},S_{k-1}),(B_1, S'),\ldots ,(B_m, S'), (A_{k+1},S_{k+1}),\ldots,(A_n,S_n)]\ \big)\] where $B_0\sim_{\theta} A_k$ for some program clause $B_0\leftarrow B_1,\ldots,B_m$ and some $k$,  $S'=S_k\cup\{A_k\}$. 

\item \emph{co-SLD loop detection ($G\rightarrow_\infty G'$):} Let $G=[(A_1,S_1),\ldots,(A_n, S_n)]$. If $A_k\approx_\theta B$ for some $k$ and some $B\in S_k$. Then we derive \[G'=  \theta\big(\ [(A_1, S_1),\ldots,(A_{k-1}, S_{k-1}),(A_{k+1}, S_{k+1}),\ldots,(A_n,S_n)]\ \big) \] 

\item \emph{restricted loop detection ($G\rightarrow_\text{co}G'$):} Let $G=[(A_1,S_1),\ldots,(A_n, S_n)]$. If $A_k\approx_\theta B$ for some $k$ and some $B\in S_k$, and $B'$ is an instance of $A_k$,  where $B'$ is a fresh-variable variant of $B$. Then we derive \[G'=  [(A_1, S_1),\ldots,(A_{k-1}, S_{k-1}),(A_{k+1}, S_{k+1}),\ldots,(A_n,S_n)]\] 
\item \emph{co-S-reduction ($G \leadsto^S_\text{co}G' $):} $G\leadsto^S_\text{co}G'$ if $G\leadsto^S G'$ or $G\rightarrow_\text{co}G'$.
\end{itemize}
\end{definition}

Note that, unlike CoLP that uses no occurs check at all, our definition of co-S-reduction still relies on occurs check within S-reductions.  
The next difference to notice is our use of the ``restricted loop detection" rule instead of the ``co-SLD loop detection" rule defined in~\cite{GuptaBMSM07,SimonBMG07,AnconaCoSLD15}.
The below examples explain the motivation behind the introduced restriction:

\begin{example}[Undesirable effect of circular unification without occurs check]\label{exmp: naive co-s-reso dead-end}
Consider the following universal and observationally productive program that resembles the one in Case 3 of Section~\ref{sec:ex}:\\
0. $ \mathtt{p(X,s(X))} \leftarrow \mathtt{q(X)}$ \\ 
1. $ \mathtt{q(s(X))}\leftarrow \mathtt{p(X,X)}$

If we use the co-SLD loop detection rule instead of the restricted loop detection rule in the definition of co-S-reduction,  we would have the following co-S-refutation:  

\noindent
$[\big(p(X,s(X)),\emptyset\big)]$ $\rightarrow$ $[\big(q(X),\{p(X,s(X))\}\big)]$ $\hookrightarrow^{X\mapsto s(X_1)}$
$[\big(q(s(X_1)),\{p(s(X_1),s^2(X_1))\}\big)]$   $\rightarrow$\\
\underline{$[\big(p(X_1,X_1),\{q(s(X_1)),p(s(X_1),s^2(X_1))\}\big)]$} $\rightarrow^{X_1\mapsto s(X_1)}_{\infty}$
$[\ ]$ 

The answer would be given by composition $\{X\mapsto s(X_1)\}\{X_1\mapsto s(X_1)\}=\{X\mapsto s^\omega,X_1\mapsto s^\omega\}$.
However, as Case 3 of Section~\ref{sec:ex} explains, this derivation in CoLP style does not correspond to any SLD-computation at infinity:
both SLD- and S-derivations fail at the underlined goal because the subgoal $p(X_1,X_1)$ does not unify with any program clause (recall that unification with occurs check is used in SLD- and S-derivations). 
The restricted loop detection rule will also fail at the underlined goal, because there is no matcher for $p(X_1,X_1)$ and $p(s(X'),s^2(X'))$.
\end{example}

\noindent The next example presents another case where our restriction to the loop detection is necessary:

\begin{example}[Ensuring precision of the answers by co-S-resolution]\label{exmp: naive co-s-reso wild instance}
Consider the following universal and observationally productive program: 

\noindent
$\mathtt{p(Y,s(X))} \leftarrow \mathtt{p(f(Y),X)}$

Again, if we use the co-SLD loop detection rule instead of the restricted loop detection rule,  we would have the following co-S-refutation:  

\noindent
$[\big(p(Y,s(X)),\emptyset\big)]$ $\rightarrow$ $[\big(p(f(Y),X),\{p(Y,s(X))\}\big)]$ $\hookrightarrow^{X\mapsto s(X_1)}$
$[\big(p(f(Y),s(X_1)),\{p(Y,s^2(X_1))\}\big)]$   $\rightarrow$
\underline{$[\big(p(f^2(Y),X_1),\{p(f(Y),s(X_1)),p(Y,s^2(X_1))\}\big)]$} $\rightarrow^{Y\mapsto f(Y),X_1\mapsto s(X_1)}_{\infty}$
$[\ ]$ 

The answer is given by composition $\{X\mapsto s(X_1)\}\{Y\mapsto f(Y),X_1\mapsto s(X_1)\}$ which 
instantiates goal $p(Y,s(X))$ to the infinite term $t_1=p(f^\omega,s^\omega)$. 

Now consider the infinite fair S-derivation for the same goal:\\
$[p(Y,s(X))]$ $\rightarrow$ $[p(f(Y),X)]$ $\hookrightarrow^{X\mapsto s(X_1)}$
$[p(f(Y),s(X_1))]$   $\rightarrow$  $[p(f^2(Y),X_1)]$  $\hookrightarrow^{X_1\mapsto s(X_2)}$ \\ $[p(f^2(Y),s(X_2))]$$\rightarrow$$[p(f^3(Y),X_2)]$$\hookrightarrow$ $\cdots$\\
By the discussion of the previous section, we can construct a corresponding infinite SLD-derivation, which computes at infinity the infinite term $t_2=p(Y,s^\omega)$. 
Thus, in this case, there is an SLD-computation at infinity that corresponds to our proof  by co-SLD loop detection rule, but it does not approximate
$t_1=p(f^\omega,s^\omega)$.

In contrast, the restricted loop detection rule will fail at the underlined goal because neither $p(Y',s^2(X'_1))$ nor $p(f(Y'),s(X'_1))$ is an instance of $p(f^2(Y),X_1)$.
 
\end{example}

The next theorem is our main result: 

\begin{theorem}[Soundness of co-S-resolution relative to SLD-computations at infinity]\label{th:main}
Let $P$ be an observationally productive and universal logic program, and $t\in\mathbf{Term}(\Sigma)$ be an atomic goal.  If there exists a co-S-refutation  
 for $P$ and $t$ that involves the restricted loop detection reduction, and 
computes the substitution $\theta$ then\\
 1) there exists an infinite fair S-derivation for $P$ and $t$, and\\ 
2) there is a term $t^\infty\in \mathbf{Term}^\infty(\Sigma)$ SLD-computed at infinity  that is a variant of $\theta(t)$.
\end{theorem} 
\begin{proof}[Proof Sketch]
The full proof is given in Appendix A.5, it develops a method of ``decircularization'', or infinite unfolding, 
of circular substitutions computed by the restricted loop detection of co-S-resolution. We then show how an infinite number of S-resolution steps corresponds to computation of the  infinite term resulting from applying decircularization. 
Similarly to related work, e.g. \protect\cite{AnconaCoSLD15}, we relate the use of loop detection to the existence of an infinite derivation. 
However, our proof does not restrict the shape of corecursion to some simple form, e.g. mutual recursion as in  \protect\cite{AnconaCoSLD15}.
This opens a possibility for future extension of this proof method.
 \end{proof}
Similarly to any other loop detection method, co-S-resolution is incomplete relative to SLD-computations at infinity. Taking any logic program that defines irregular streams (cf. Example~\ref{ex:fibs})  will result in S-derivations with loops where subgoals do not unify.

Because the restricted loop detection is an instance of the co-SLD loop detection, and because there is a one-to-one correspondence between SLD- and S-resolution reductions, we can prove that co-S-resolution is sound relative to the greatest complete Herbrand models, by adapting the proof of~\cite{GuptaBMSM07,SimonBMG07,AnconaCoSLD15}, see Appendix A.4.


\section{Conclusions, Discussion, Related and Future Work}\label{sec:conclusion}

\paragraph{Conclusions} We have given a computational characterisation to  the  \emph{SLD-computations at infinity}~\cite{Llo88,EmdenA85}
introduced in the 1980s. 
Relying on the recently proposed notion of observational productivity of logic programs, we have shown that infinite observationally productive derivations are
  sound and complete  relative to the SLD-computations at infinity.
	This paper thus confirmed the conjecture made in~\cite{KJS17} that the weaker notion of observational productivity of logic programs implies the much stronger notion of global productivity of individual derivations.
	 This result only holds on extra condition -- universality of logic programs in question. This fact has not been known prior to this paper.

We have introduced co-S-resolution that gives the first algorithmic characterization of the SLD-computations at infinity. We proved that co-S-resolution is sound relative to the SLD-computations at infinity for universal and observationally productive programs.
Appendix A.6 discusses its implementation.
 Structural resolution, seen as a method of systematic separation of SLD-resolution to mgu and mgm steps, has played an instrumental role
in the proofs.

\paragraph{Discussion} Imposing the conditions of observational productivity and universality allowed us to study the operational properties of infinite productive SLD-derivations without introducing any additional 
modifications to the resolution algorithm.  These results can be directly applied in any already existing Prolog implementation: ensuring that a program satisfies these conditions ensures that all infinite computations for it are productive.   The results can similarly  be reused in any other inference algorithm based on resolution; this paper showed its adaptation to CoLP. 

As a trade-off, both conditions exclude logic programs that may give rise to globally productive SLD derivations for certain queries.
For example, a non-productive logic program that joins clauses of examples~\ref{ex:natstream} and~\ref{ex:bad} will still result 
in productive derivations for a query $\mathtt{nats(X)}$, as computations calling the $\mathtt{bad}$ clause would not interfere in such derivations. 
The universality condition excludes two cases of globally productive derivations:

(1) Existential variables in the clause body have no effect on the arguments in which the infinite term is computed. E.g., taking a coinductive definition
$\mathtt{p(s(X),Y) \gets p(X,Z)}$ and a goal  $\mathtt{p(X,Y)}$, an infinite term will be produced in the first argument, and the existential variable occurring in the second argument plays no role in this computation.

(2) Existential variables play a role in SLD-computations of infinite terms at infinity. 
In such cases, they usually  depend on other variables in the coinductive definition, and their productive instantiation is  guaranteed by other clauses in the program. 

The famous  definition of the stream of Fibonacci numbers is an example: 


\begin{example}[Fibonacci Numbers]\label{ex:fibs}
\noindent
0. $\mathtt{add(0,Y,Y)\gets}$\\ 
1. $\mathtt{add(s(X),Y,s(Z)) \gets add(X,Y,Z)}$\\
2. $\mathtt{fibs(X,Y,[X|S]) \gets add(X,Y,Z),fibs(Y,Z,S)}$

\noindent The goal $\mathtt{fibs(0,s(0),F)}$ computes the infinite substitution  $\mathtt{F\mapsto [0,s(0),s(0),s^2(0),\ldots]}$ at infinity. 
The existential variable $\mathtt{Z}$  in the body of clause (2.) is instantiated by the $\mathtt{add}$ predicate, as $Z$ in fact functionally depends on $X$ and $Y$ (all three variables contribute to construction of the infinite term in the third argument of $\mathtt{fibs}$). 
\end{example}

\noindent In the future, these classes of programs may be admitted by following either of the two directions:

(I) Using program transformation methods to ensure that every logic program is transformed into observationally productive and universal form. An example of the observationally productive transformation is given by~\cite{FK16}, and of the universal transformation -- by~\cite{EVarElim_for_CLP_Senni2008,exists_Var_Elim_for_LP_PROIETTI95}. This approach may have drawbacks, such as changing the coinductive models of the programs.
  
(II) Refining the resolution algorithm and/or the loop detection method. 
 For example, the restrictions imposed on the loop detection in Definition~\ref{df:co-S} can be further refined. Another solution is to 
use the notion of \emph{local productivity}~\cite{FK16} instead of the observational productivity. A derivation is locally productive if it computes an infinite term only at a certain argument. Both examples given in this section are locally productive.
In \cite{FK16}, local productivity required technical modifications to the unification algorithm (involving labelling of variables), and establishing its soundness is still an open problem.

\paragraph{Related Work} The related paper~\cite{Y17} shows that an algorithm embedding the CoLP loop detection rule into S-resolution is sound relative to greatest complete Herbrand models. However, that work does not consider conditions on which such embedding would be sound relative to SLD-computations at infinity. In fact, as this paper shows, simply embedding the CoLP loop detection rule into S-resolution does not make the resulting coinductive proofs
sound relative to SLD-computations at infinity, as the three undesirable cases of Section~\ref{sec:ex} may still  occur.   
The construction of the proof of Theorem~\protect\ref{th:main} in this paper agrees with a similar construction for non-terminating SLDNF derivation \protect\cite{Shen2003InfiniteSLDNF}. We show that the use of restricted loop detection corresponds to infinite S-derivation characterized by infinitely repeating subgoal variants, while \protect\cite{Shen2003InfiniteSLDNF} shows that a non-terminating SLDNF derivation admits an infinite sequence of subgoals that are either variants or increasing in size.
 
\paragraph{Future Work} 
Similarly to other existing loop detection methods, co-S-resolution is incomplete, as it only captures regular infinite terms. Future work will be to introduce heuristics 
extending our methods to irregular structures.
Another  direction for future work is to investigate practical applications of co-S-resolution in Internet programming and type inference in programming languages, as was done in~\cite{FKSP16}. 

\bibliographystyle{acmtrans}
\bibliography{katya2}

\vfill

\pagebreak
\appendix
\section{Supplementary Materials and Full Proofs}\label{sec:app}

\subsection{Least and Greatest Complete Herbrand Models}\label{sec:app2}

We recall the least and greatest complete Herbrand model constructions
for LP~\cite{Llo88}. We express the definitions in the form of a
big-step semantics for LP, thereby exposing duality of inductive and
coinductive semantics for LP in the style of~\cite{Sg12}. We start
by giving inductive interpretations to logic programs. We say that $\sigma$
is a {\em grounding substitution for $t$} if $\sigma(t) \in
\textbf{GTerm}^{\omega}(\Sigma)$, and is just a {\em ground
  substitution} if its codomain is $\textbf{GTerm}^{\omega}(\Sigma)$.

\begin{definition}\label{df:irules}
Let $P$ be a logic program. The {\em big-step rule for $P$} is given
by
\[\frac{P \models \sigma(B_1), \,\ldots \, ,P \models \sigma(B_n)}{P
  \models \sigma(A)}\] where $A \gets B_1, \ldots B_n$ is a clause in
$P$ and $\sigma$
is a grounding
substitution.
\end{definition}

\noindent
Following standard terminology \cite{ACZEL1977InroIndc,Sg12},
 we say that
an inference rule is \emph{applied forward} if it is applied from top
to bottom, and that it is \emph{applied backward} if it is applied
from bottom to top. If a set of terms is closed under forward
(backward) application of an inference rule, we say that it is {\em
  closed forward} (resp., {\em closed backward}) under that rule.  If
the $i^{th}$ clause of $P$ is involved in an
application of the big-step rule for $P$, then we may say that we have
applied the {\em big-step rule for $P(i)$}.

\begin{definition}\label{def:model}
The {\em least Herbrand model} for a program $P$ is the
smallest set $M_P \subseteq \cat{GTerm}(\Sigma)$ that is closed
forward under the big-step rule for $P$.
\end{definition}
\begin{example}
The least Herbrand model for the program of Example $1.1$ is 
$\{\mathtt{nat(0)},$ $\mathtt{nat(s(0))},$ $\mathtt{nat(s^2(0)},
\ldots\}$. We use $\mathtt{s^2}(0)$ for $\mathtt{s(s(0))}$, $s^3(0)$ for $\mathtt{s(s(s(0)))}$ and so on.
\end{example}

The requirement that $M_P\subseteq \cat{GTerm}(\Sigma)$ entails that
only ground substitutions are used in
the forward applications of the big-step rule involved in the
construction of $M_P$. Next we give coinductive interpretations to
logic programs. For this we do not impose any finiteness requirement
on the codomain terms of $\sigma$.

\begin{definition}\label{def:cmodel}
The {\em greatest complete Herbrand model} for a program $P$ is the largest set $M^{\omega}_P \subseteq
\cat{GTerm}^{\omega}(\Sigma)$ that is closed backward under the
big-step rule for $P$.
\end{definition}

\begin{example}[Complete Herbrand model]
The greatest complete Herbrand model for the program of Example $1.1$ is $\{\mathtt{nat(0)},$
$\mathtt{nat(s(0))},$ $\mathtt{nat(s^2(0))}$ $\ldots\}\bigcup
\{\mathtt{nat(s^\omega)}\}$.  Indeed, there is an infinite inference
for $\mathtt{nat(s^\omega)}=\mathtt{nat(s(s(...)))}$ obtained by repeatedly applying the
big-step rule for this program backward.
\end{example}

Definitions~\ref{def:model} and~\ref{def:cmodel} could alternatively
be given in terms of least and greatest fixed point operators, as in,
e.g.,~\cite{Llo88}. To ensure that $\cat{GTerm}(\Sigma)$ and
$\cat{GTerm}^{\omega}(\Sigma)$ are non-empty, and thus that the least
and greatest Herbrand model constructions are as intended, it is
standard in the literature to assume that $\Sigma$ contains at least
one function symbol of arity $0$.
We will make this assumption throughout the remainder of this paper.

\subsection{Proof of Productivity Lemma 4.1}\label{ap:pl}

\emph{Let $P $ be an observationally productive and universal program 
and   let $t \in
\mathbf{Term}(\Sigma)$. 
Let $D$ be an infinite S-resolution derivation given by $G_0=t \leadsto^S G_1 \leadsto^S G_2 \leadsto^S \ldots$, then
 for every $G_i \in D$, there is a $G_j \in D$, with $j > i$, such that, given computed mgus $\theta_i, \ldots, \theta_1$ up to $G_i$ and the computed mgus $\theta_j, \ldots, \theta_1$ up to $G_j$, $d(t^\infty,\theta_i, \ldots, \theta_1(t)) > d(t^\infty,\theta_j, \ldots, \theta_1(t))$,
for some term $t^\infty \in \mathbf{Term}^{\infty}(\Sigma)$.}

The proof has two parts, as follows. Part 1 shows that, under the imposed productivity and universality conditions, no infinite sequence of trivial unifiers is possible for infinite S-resolution derivations. Therefore, an infinite S-resolution derivation must contain an infinite number of non-trivial substitutions.  Part 2 uses this fact and shows that a composition of an infinite number of non-trivial substitutions must result in an infinite term (this holds under universality condition only). 

\begin{proof}

Recall that,
by definition of S-resolution reductions, each step $G_k \leadsto^S G_{k+1}$ is a combination of a finite number of steps $G_k \rightarrow^n [A_1, \ldots,  A_n]$  and one substitution+rewriting step $ [A_1, \ldots, A_j, \ldots ,  A_n] \rightarrow \circ \hookrightarrow G_{k+1}$, this final step involves computation of an mgu (but not mgm) $\theta_{k+1}$ of some clause $C \gets C_1, \ldots, C_n$ and some $A_j$. 
So in fact \[G_{k+1} = [\theta_{k+1}(A_1), \ldots, \theta_{k+1}(A_{j-1}), \theta_{k+1}(C_1), \ldots, \theta_{k+1}(C_n), \theta_{k+1}(A_{j+1}), \ldots,  \theta_{k+1}(A_n)]\]
Moreover, since $\theta_{k+1}$ is not an mgm, we have that:\\
\emph{$\theta_{k+1}$ is a non-trivial substitution for at least one variable $X$ in $A_j$.  \ \ \ \ (**)}

We will use the above facts implicitly in the  proof below.



To proceed with our proof, first we need to show that 

(1) For all $k>1$ in $G_0=t \leadsto^S G_1 \leadsto^S G_2 \leadsto^S \ldots \leadsto^S G_k \leadsto^S \ldots$,  the composition $\theta_{k} \ldots \theta_1$ 
is non-trivial for $t$. 

We prove this by induction. 

\emph{Base case.} By $(**)$,  $\theta_1$ is necessarily non-trivial for initial goal $t$.\\
\emph{Inductive case.} If $\theta_k$ is non-trivial for term $\theta_{k-1} \ldots \theta_1(t)$, then by universality of $P$ and $(**)$, $\theta_{k+1}$ is non-trivial for term $\theta_{k} \ldots \theta_1(t)$. Then by induction, for all $k>1$,  the composition $\theta_{k} \ldots \theta_1$ is non-trivial for $t$.

Next, we need to show that the property (1) implies that\\

(2) we can define the limit term $t^\infty$ using the infinite sequence \[t,\quad\theta_1(t),\quad\theta_2\theta_1(t),\quad\theta_3\theta_2\theta_1(t),\quad\ldots\] 

To prove (2), we prove the following property:

(2.1) For each  $n \in \Nat$, there exists  $\theta_{k_n}$, so that for all $k$, if $k>k_n$, then truncation of $\theta_k\ldots\theta_1(t)$ at depth $n$ is the same as the truncation of $\theta_{k_n}\ldots\theta_1(t)$ at depth $n$. 

We prove this fact by contradiction. Assume the negation of our proposition, which says there exists depth value n such that for all substitution subscript $k$, there exists some $k_n$, so that $k_n>k$ and truncation of $\theta_{k_n}\ldots\theta_1(t)$ at depth n is different from the truncation of $\theta_{k}\ldots\theta_1(t)$ at depth n. This is impossible because this implies that non-trivial substitution can be infinitely applied within the truncation at depth n but no finitely branching tree can accommodate infinite amount of variables up to any fixed depth.

This gives us a way to prove (2):

We build $t^\infty$ inductively, for each depth $n$ of $t^\infty$.
For depth $n= 0$, we let $t^\infty$ have as its root symbol the predicate symbol $t(\epsilon)$ of initial atomic goal $t$.  If $t^\infty$ is defined up till depth $n\geq0$, then, by (2.1) we know that there is some $k_n$ such that for all $k>k_n$, 
\[\gamma\:'(n,\theta_{k}\ldots\theta_1(t))=\gamma\:'(n,\theta_{k_n}\ldots\theta_1(t))\]
 
We also know by (2.1) that there is some $k_{n+1}$ such that for all $k>k_{n+1}$, \[\gamma\:'((n+1),\theta_{k}\ldots\theta_1(t))=\gamma\:'((n+1),\theta_{k_{n+1}}\ldots\theta_1(t))\] Then we find the greater value $\kappa$ in $\{k_n,k_{n+1}\}$, or set $\kappa=k_n$ if $k_n=k_{n+1}$, and define the nodes at depth $n+1$ for $t^\infty$ in the same way as $\theta_\kappa\ldots\theta_1(t)$.



\end{proof}

\subsection{Proof of Theorem 4.1 Soundness and Completeness of Infinite S-resolution Relative to SLD-computations at Infinity}\label{ap:sSr}
\emph{Let $P $ be an observationally productive and universal program, and 
  let $t \in
\mathbf{Term}(\Sigma)$.
There is an infinite fair S-resolution derivation for $t$ iff there is a $t' \in
\cat{Term}^\infty(\Sigma)$, such that $t'$ is SLD-computable at infinity by $t$.}

Proofs in both directions start with establishing operational equivalence of \emph{infinite} S-resolution and SLD-resolution derivations.
Coinductive proof principle is employed in this part of the proof. The proof in the left-to-right direction proceeds by using this equivalence, and by applying Lemma 4.1 to show that an infinite fair S-resolution derivation must result in an SLD-computation of an infinite term at infinity. 
The other direction is proven trivially from the operational equivalence of infinite S-resolution and SLD-resolution derivations.

\begin{proof}

\begin{enumerate}
	\item Suppose $D = t \leadsto^S G_1 \leadsto^S G_2 \leadsto^S \ldots$ is an infinite fair S-derivation.
It is easy to construct a corresponding SLD-resolution derivation $D^*$, we prove this fact by coinduction.
Consider $t \leadsto^S G_1$, which in fact can be given by one of two cases:
\begin{enumerate}
	\item $t \hookrightarrow \theta(t) \rightarrow G_1$, i.e. if $t$ does not match, but is unifiable with some clause $P(i)$ via a substitution $\theta$. 
	In this case, the first step in $D^*$ will be to apply SLD-resolution reduction to $t$ and $P(i)$: $t \leadsto G_1$. 
	\item $t \rightarrow^n [A_1, \ldots, A_j, \ldots , A_n] \hookrightarrow [\theta(A_1), \ldots, \theta(A_j), \ldots , \theta(A_n)] \rightarrow G_1$;  obtained by resolving $A_j$ with a clause $P(i)$ and computing a substitution $\theta$. Then, in $D^*$, we will have $n$ steps by SLD-resolution reductions involving exactly the resolvents of goal atoms and clauses used in $t \rightarrow^n [A_1, \ldots , A_n]$ (note that mgms used in $\rightarrow^n$ are also mgus by definition). These $n$ steps in $D^*$ will be followed by one step of SLD-resolution reduction, resolving $A_j$ with $P(i)$ using substitution $\theta$. 

We can proceed coinductively to construct $D^*$ from $D$ starting with $G_1 \in D$.
\end{enumerate}

We now need to show that such $D^*$ is fair and non-failing.
By definition, $t \leadsto^S G_1 \leadsto^S G_2 \leadsto^S \ldots$ should contain atoms which are resolved against
finitely often. This means that corresponding derivation $D^*$ will be fair.
Because $D$ is non-terminating and non-failing, $D^*$ using the same resolvents will be non-terminating and non-failing, too. 

Finally, we need to show that $D^* = t \leadsto G^*_1 \leadsto G^*_2 \leadsto \ldots$ constructed as described above involves computation of an infinite term $t'$ at infinity. This can only happen if, 
for every $G^*_i \in D^*$, there is a $G^*_j \in D^*$, with $j > i$, such that, given computed mgus $\theta_i, \ldots, \theta_1$ up to $G^*_i$ and the computed mgus $\theta_j, \ldots, \theta_1$ up to $G^*_j$, $d(t',\theta_i, \ldots, \theta_1(t)) > d(t',\theta_j, \ldots, \theta_1(t))$.
For this to hold, the S-resolution derivation $D$ should satisfy the same property, but this follows from Lemma 4.1. 

\item 
The proof proceeds by coinduction.
Consider the SLD-resolution derivation $D^* = t \leadsto G^*_1 \leadsto G^*_2 \leadsto \ldots $ that computes an infinite term $t'$ at infinity.
Consider the substitution $\theta$ associated with $t \leadsto G^*_1$. If it is an mgm of $t$ and some clause $P(i)$, then we can construct the first step of S-resolution reduction  using the rewriting reduction:  $ t \rightarrow G^*_1$.  If $\theta$ is not an mgm, i.e. it is an mgu, then we can construct first two steps of the S-resolution reduction: $ t \hookrightarrow \theta(t) \rightarrow G^*_1$. We can proceed building $D$ from $D^*$ in the same way, now starting from $G^*_1$. We only need to show that $D$ is fair and non-failing, but that follows trivially from properties of $D^*$. 

\end{enumerate}
\end{proof}

\subsection{Standard co-SLD-resolution and Proof of Soundness of Co-S-resolution}\label{ap:cosld}


In this subsection, we introduce the standard definition of co-SLD-derivations~\cite{AnconaCoSLD15}, and re-use the proof of their soundness with respect to the greatest complete Herbrand models to establish a similar result for co-S-resolution.

\begin{definition}[Co-SLD-reductions~\cite{AnconaCoSLD15}]\label{defn: co-s-reso}
Given a logic program P, we distinguish the following reductions in the context of co-inductive logic programming.
\begin{itemize}
\item \emph{SLD reduction ($G\leadsto G'$):} Let $G=[(A_1,S_1),\ldots,(A_n, S_n)]$. If $B_0\approx_{\theta} A_k$ for some program clause $B_0\leftarrow B_1,\ldots,B_m$ and some $k$, then let $S'=S_k\cup\{A_k\}$, we derive  \[G'=\theta \big(\ [ (A_1, S_1),\ldots,(A_{k-1},S_{k-1}),(B_1, S'),\ldots ,(B_m, S'), (A_{k+1},S_{k+1}),\ldots,(A_n,S_n)]\ \big)\]

\item \emph{loop detection ($G\rightarrow_\infty G'$):} Let $G=[(A_1,S_1),\ldots,(A_n, S_n)]$. If $A_k\approx_\theta B$ for some $k$ and some $B\in S_k$, we derive \[G'=  \theta\big(\ [(A_1, S_1),\ldots,(A_{k-1}, S_{k-1}),(A_{k+1}, S_{k+1}),\ldots,(A_n,S_n)]\ \big)\] 

\item \emph{co-SLD reduction ($G\leadsto_\text{co} G'$):} $G\leadsto_\text{co} G'$ if $G\leadsto G'$ or  $G\rightarrow_\infty G'$.

\end{itemize}
\end{definition}

Co-SLD-resolution is proven sound in~\cite{AnconaCoSLD15,SimonCoSLD2006}, i.e. if a logic program $P$ and an atomic  goal $G$ have a co-SLD-refutation with computed answer substitution $\theta$, then all ground instances of $\theta(G)$ are in the greatest complete Herbrand model of $P$. 


An important property of  co-S-resolution is coinductive soundness. 
\begin{proposition}[Soundness of co-S-resolution]\label{lem: soundness of naive co-S reso}
If a logic program P and an atomic initial goal G have a co-S-refutation with computed answer substitution $\theta$, then all ground instances of $\theta(G)$ are in the greatest complete Herbrand model of P. 
\end{proposition} 
 We will base the proof on the soundness of co-SLD resolution~\cite{AnconaCoSLD15,SimonCoSLD2006}.   
\begin{proof}
If loop detection is not used at all in the co-S-refutation, then the co-S-refutation reduces to a S-refutation, which is sound w.r.t to least Herbrand model, thus also being sound w.r.t the greatest complete Herbrand model.

Let us assume loop detection is used for at least once. We show that for any co-S-refutation there exists a corresponding co-SLD refutation. Any substitution+rewriting step $G_i\hookrightarrow G_{i+1}\rightarrow G_{i+2}$ corresponds to one step of SLD reduction (in co-SLD setting) $G_i\leadsto G_{i+2}$. Any rewriting reduction step $G_i\rightarrow G_{i+1}$ that does not follow a substitution reduction step also constitutes a SLD-reduction step $G_i\leadsto G_{i+1}$. In this way any co-S-refutation can be converted to a refutation that only involves SLD-reduction and loop detection, thus constituting a co-SLD refutation, which is sound w.r.t the greatest complete Herbrand model. 
\end{proof}

\subsection{Proof of Theorem 5.1 of Soundness of Co-S-resolution Relative to SLD-Computations at Infinity}\label{ap:mainth}

\emph{
Let $P$ be an observationally productive and universal logic program, and $t\in\mathbf{Term}(\Sigma)$ be an atomic goal.  If there exists a co-S-refutation  
 for $P$ and $t$ that involves the restricted loop detection rule, and computes the substitution $\theta$ then}

  \begin{enumerate}
  \item \emph{there exists an infinite fair S-derivation for $P$ and $t$, and}
  \item \emph{there is a term $t^\infty\in \mathbf{Term}^\infty(\Sigma)$ SLD-computed at infinity  that is a variant of $\theta(t)$.}
  \end{enumerate}

The proof will proceed according to the following scheme.
 For the sake of the argument, we take some arbitrary logic program that satisfies the productivity and universality conditions. We first show that the use of (any) loop detection necessarily results in computation of circular substitutions. Next, we analyse the effect of
	the restriction that was introduced to loop detection in Definition 5.1 and  
	build the infinite regular S-derivation starting at the point where the restricted loop detection was once used. Finally, we show that the sequence of (non-circular) unifiers computed by the infinite S-derivation is equivalent to the single (circular) unifier computed by the restricted loop detection of Definition 5.1. If there are several uses of the restricted loop detection rule, then each implies a separate infinite derivation, and they can be interleaved to form an infinite fair S-derivation. This argument relies only on the observational productivity, leading to the conclusion  that \emph{for an observationally productive program, if it has a co-S-refutation
	then there exists an infinite fair S-derivation in which a sequence of computed unifiers ``unfolds" the circular unifier}. A program that is also universal is a special case, where (by Theorem 4.1) the infinite sequence of unifiers 
	instantiates the initial goal into an infinite formula, which shall be a variant of the formula computed by co-S-refutation. 

Before proceeding with the full proof, we first need to introduce the method of \emph{decircularization}.
A circular substitution means a substitution of infinite regular terms for variables. For e.g. $\{X\mapsto s(X)\}$ is equivalent to $\{X\mapsto s^\omega\}$, where $s^\omega$ is obtained by continued substitution of $s(X)$ for $X$, which can further be regarded as applying an infinite succession of non-circular substitutions $s(X_1)$ for $X$, $s(X_2)$ for $X_1$, $s(X_3)$ for $X_2$, and so on.  We coin the term  \emph{decircularization} for the process of obtaining from a circular substitution $\sigma$ an equivalent infinite succession of non-circular substitutions $\sigma_1,\sigma_2,\ldots$. In the following proof we relate co-S-resolution's answers to terms SLD-computable at infinity by showing that the circular substitutions computed by co-S-refutation have decircularization computed by infinite S-derivation. Formal definition of decircularization (with motivating examples) is given below.

\begin{definition}[Decircularization]
Let  $\sigma=\{\ldots,X_k\mapsto t,\ldots\}$ be a circular substitution where $X_k\mapsto t$ a circular component and $FV(t)=\{X_1,\ldots,X_k,\ldots,X_m\}$.  $X_k\mapsto t$ can be \emph{decircularized} into  an infinite set  $R=\{X_{k}\mapsto t_{(1)}, X_{k_{(1)}}\mapsto t_{(2)}, X_{k_{(2)}}\mapsto t_{(3)},\ldots\} $ where $t_{(n)}$ is a variant of $t$ obtained by applying renaming $\{X_i\mapsto X_{i_{(n)}}\mid \forall i\in[1,m]  \}$ to $t$. We call the set $R$ a \emph{decircularization} of $X_k\mapsto t$. The \emph{decircularization} of $\sigma$ is the union of all decircularizations of individual circular components of $\sigma$. 
\end{definition}

\begin{example}[Decircularization]\label{exmp: decircularization}
Let $\sigma=\{A_1\mapsto f(A_1,B_1,C_1),B_1\mapsto s(B_1)\}$. The decircularization of $\sigma$ is $R_A\cup R_B$ where  $R_A=\{A_1\mapsto f(A_{1_{(1)}},B_{1_{(1)}},C_{1_{(1)}}),A_{1_{(1)}}\mapsto f(A_{1_{(2)}},B_{1_{(2)}},C_{1_{(2)}}),A_{1_{(2)}}\mapsto f(A_{1_{(3)}},B_{1_{(3)}},C_{1_{(3)}}),\ldots\}.$ and $R_B=\{B_1\mapsto s(B_{1_{(1)}}),B_{1_{(1)}}\mapsto s(B_{1_{(2)}}),B_{1_{(2)}}\mapsto s(B_{1_{(3)}}),\ldots\}.$ With the understanding that subscriptions merely serve the purpose of distinguishing names, we can simplify the decircularization into  $R_A=\{A_1\mapsto f(A_2,B_2,C_2),A_2\mapsto f(A_3,B_3,C_3),A_3\mapsto f(A_4,B_4,C_4),\ldots\}$ and $R_B=\{B_1\mapsto s(B_2),B_2\mapsto s(B_3), B_3\mapsto s(B_4),\ldots\}$. Note the way A's and B's interact in the decircularization.

\end{example}  
\begin{example}\label{exmp: working case of naive co-s--reso}
Consider program:

\noindent
r(f(A,B,C), s(B)) $\gets$ r(A,B).

A co-S-refutation $D$ is: 

\noindent
$[\big(r(X,Y),\emptyset\big)]$ $\hookrightarrow^{X\mapsto f(A_1,B_1,C_1),Y\mapsto s(B_1)}$ $[\big(r(f(A_1,B_1,C_1),s(B_1)),\emptyset\big)]$ $\rightarrow$\\ \underline{$[\big(r(A_1,B_1),\{r(f(A_1,B_1,C_1),s(B_1))\}\big)]$} $\rightarrow^{A_1\mapsto f(A_1,B_1,C_1), B_1\mapsto s(B_1)}_{\text{co}}$ $[\ ] $ 

\vspace*{0.08in}
If we continue the derivation from the underlined goal in $D$, but now use S-resolution, we have an infinite S-derivation $D^*$ as follows:

\noindent
$[r(X,Y)]$ $\hookrightarrow^{X\mapsto f(A_1,B_1,C_1),Y\mapsto s(B_1)}$ $[r(f(A_1,B_1,C_1),s(B_1))]$ $\rightarrow$ \\ \underline{$[r(A_1,B_1)]$} $\hookrightarrow^{A_1\mapsto f(A_2,B_2,C_2),B_1\mapsto s(B_2)}$ $[r(f(A_2,B_2,C_2),s(B_2))]$ $\rightarrow$ \\$[r(A_2,B_2)]$ $\hookrightarrow^{A_2\mapsto f(A_3,B_3,C_3),B_2\mapsto s(B_3)}$ $[r(f(A_3,B_3,C_3),s(B_3))]$ $\rightarrow$\\ $[r(A_3,B_3)]$ $\cdots$

Note that the mgu's computed by infinite S-derivation starting from the underlined goal  in $D^*$ is a decircularization of the circular substitution computed by co-S-resolution from the corresponding goal in $D$. The details of computing decircularization for the circular substitution of this example is given in Example~\ref{exmp: decircularization}.  We see that in this example co-S-resolution is a perfect finite model for corresponding infinite S-resolution.
\end{example}

\begin{proof}
We first take an arbitrary coinductive logic program that satisfies our productivity and universality conditions.     
\begin{enumerate}

\item \emph{We first show that the use of loop detection (no matter restricted or not) necessarily results in creation of circular substitutions.}

 Generally, consider some subgoal of the form $(A,\{A_1,\ldots A_n\})$ where for all $A_i$ in the ancestors set of $A$, $A_i$ is added to the ancestors set later than $A_{i+1}$. By definition of co-S-derivation, there exists some program clause instances (or  variants)

\begin{tabular}{l}
$A_n \leftarrow \ldots,A_{n-1},\ldots$\\
$\vdots$\\
$A_2 \leftarrow \ldots,A_1,\ldots$\\
$A_1 \leftarrow \ldots,A,\ldots$\\
\end{tabular}
where all $A$ and $A_k\ (1\leq k\leq n)$ are finite, so are the omitted atoms (which are represented by ``$\ldots$'' ) in the above set of clauses.

Assume \emph{restricted loop detection} is applicable for the subgoal under consideration.  This means  that $A$ unifies with some $A_i$ (occurs check switched off) under mgu $\sigma$,  and $A\prec A_i'$ where $A_i'$ is a fresh-variable variant of $A_i$.  Each of the two conditions $A\approx_\sigma A_i$ and $A\prec A_i'$ has implication.

Note that $\sigma$ is a circular substitution: if $\sigma$ is not a circular substitution, then we have the set of program clause instances (or  variants)

\begin{tabular}{l}
$\sigma(A_i) \leftarrow \ldots,\sigma(A_{i-1}),\ldots$\\
$\vdots$\\
$\sigma(A_2) \leftarrow \ldots,\sigma(A_1),\ldots$\\
$\sigma(A_1) \leftarrow \ldots,\sigma(A),\ldots$\\
\end{tabular}
 where $\sigma(A)=\sigma(A_i)$ (because $A\approx_\sigma A_i$) and all atoms are finite. Then there exists non-terminating rewriting reduction steps for $\sigma(A_i)$ and thus breaks the observational productivity condition. Therefore $\sigma$ is circular.

\item \emph{Next we construct the general form of the repeating derivation pattern by 
analysing the effect of the restricted loop detection.  This pattern is then used to build the infinite regular derivation that can start at the point where the restricted loop detection was once used.}

An infinite sequence $\{\gamma_{n}\}_{n\geq 0}$ of trivial substitutions is defined as follows. Let $\gamma_0$ be the empty substitution and $\gamma_1$ be a variable renaming substitution for $A_i$ with fresh names.  Let $\{\gamma_{n}\}_{n\geq 1}$ be an infinite sequence of renaming substitutions, such that, for all $n\geq 2$, the domain of $\gamma_{n}$ equals to the image of $\gamma_{n-1}$, while the image of $\gamma_{n}$ is disjoint from the set of all variables that occur in the domain of one of $\gamma_{k}\ (1\leq k\leq n)$. For example, if $\gamma_1=\{X_1\mapsto X_2\}$, then the sequence of renaming substitutions $\{X_1\mapsto X_2\}, \{X_2\mapsto X_3\},\{X_3\mapsto X_4\},\ldots$ conforms to the above description for $\{\gamma_{n}\}_{n\geq 1}$.   

For all $n\geq 0$, $\gamma_{n+1}\cdots\gamma_0(A_i)$ is an instance of  $\gamma_n\cdots\gamma_0(A)$, as is implied by $A\prec A_i'$. Let $\sigma_{n+1}$ denote the matcher for the pair  $\gamma_n\cdots\gamma_0(A)$ and $ \gamma_{n+1}\cdots\gamma_0(A_i)$, we have the important equation  \[\sigma_{n+1}\gamma_n\ldots\gamma_0(A)=\gamma_{n+1}\ldots\gamma_0(A_i),\quad \text{for all } n\geq 0\] We also have a set of program clause instances (or variants) for all $n\geq 0$:

\begin{tabular}{l}
$\gamma_{n+1}\ldots\gamma_0(A_i) \quad\leftarrow\quad \ldots,\quad\gamma_{n+1}\ldots\gamma_0(A_{i-1}),\quad\ldots$\\
$\vdots$\\
$\gamma_{n+1}\ldots\gamma_0(A_2) \quad\leftarrow\quad \ldots,\quad\gamma_{n+1}\ldots\gamma_0(A_1),\quad\ldots$\\
$\gamma_{n+1}\ldots\gamma_0(A_1) \quad\leftarrow\quad \ldots,\quad\gamma_{n+1}\ldots\gamma_0(A),\quad\ldots$\\
\end{tabular}
Using the above clauses for rewriting reductions, we have that for all $n\geq 0$, there exists the repeating S-derivation \emph{pattern}:  \[[\ldots,\ \gamma_n\cdots\gamma_0(A),\ \ldots]\hookrightarrow [\ldots,\ \sigma_{n+1}\gamma_n\ldots\gamma_0(A), \ \ldots]\rightarrow^i [\ldots,\  \gamma_{n+1}\ldots\gamma_0(A),\ \ldots] \]

Therefore an infinite S-derivation starting from goal $[\ldots,A,\ldots]$ can be given in the form 

\noindent
$[\ldots,\quad \gamma_0(A),\quad \ldots]$ $\hookrightarrow$ $[\ldots,\quad\sigma_{1}\gamma_0(A),\quad\ldots]$ $\rightarrow^i$ \\
$[\ldots,\quad\gamma_{1}\gamma_0(A),\quad\ldots]$ $ \hookrightarrow$ $[\ldots, \quad\sigma_{2}\gamma_1\gamma_0(A),\quad\ldots]$ $ \rightarrow^i $\\
$[\ldots,\quad\gamma_{2}\gamma_{1}\gamma_0(A),\quad\ldots]$ $ \hookrightarrow\quad \cdots$

\item  \emph{Finally we show that the collection of (non-circular) unifiers computed by the infinite derivation is equivalent to the single (circular) unifier computed by restricted loop detection.}

Consider a circular component $X\mapsto t[X]$ of $\sigma$. This circular component corresponds to a mapping $X_n\mapsto t[X_{n+1}]$ in each matcher $\sigma_{n+1}\ (n\geq 0)$ . Therefore the collection of all such $X_n\mapsto t[X_{n+1}]$ constitutes a decircularization of $X\mapsto t[X]$, and all other circular components of $\sigma$ are similarly decircularized. Therefore $\sigma$ has the decircularization $\bigcup_{n=1}^\infty\sigma_{n}$.
\end{enumerate}

If there are several use of restricted loop detection, then each implies a separate infinite derivation, which can be interleaved to form an infinite fair S-derivation.  

So far only the observational productivity condition has been used, and the conclusion is that

\emph{For observationally productive programs, if there is a co-S-refutation involving restricted loop detection and a circular unifier $\theta$, then 1) there exists an infinite fair co-S-derivation that 2) computes an infinite sequence of unifiers equivalent to the circular 
unifier $\theta$.} \hfill ($\star$)

A program that is also universal is a special case of ($\star$), which, by 1) of ($\star$), has an infinite fair S-derivation whose unifiers, instantiate the initial goal into an infinite formula (by Theorem 4.1). But then by 2) of ($\star$), a composition of these unifiers must compute a variant of the formula computed by co-S-refutation.
\end{proof}

\subsection{Implementation of Co-S-resolution}\label{ap:impl}

The co-S-resolution meta-interpreter is written in SWI-Prolog, and is available at\\ \url{https://github.com/coalp/Productive-Corecursion}. It adopts left first computation rule and depth first search rule in the SLD tree. 

The entry procedure requires a unary predicate named \texttt{clause\_tree}, which takes a conjunctive goal or an atomic goal as an input. After assignment of an empty ancestors set to the goal, a case analysis on the shape of the goal passes an atomic goal to procedures corresponding to reduction rules, or disassembles a conjunctive goal into its head and tail, and processes the head and the tail separately and recursively, starting with a case analysis on their shape. 
   
Three kinds of reduction rule: rewriting reduction, substitution reduction and restricted loop detection are coded separately as three alternative procedures to process an atomic goal. Since object programs to be processed by the meta-interpreter are intended  to be non-terminating, and given the execution model of Prolog, it is necessary to put the loop detection rule ahead of other rules, otherwise in a non-terminating derivation it will never be called. The rewriting reduction is put at the second place, and the substitution reduction is tried only when both the loop detection and the rewriting reduction are not applicable. The ordering of the rules, therefore,  also makes sure that rewriting happens after each substitution reduction, because if a sub-goal cannot be reduced by loop detection, an instance of this sub-goal from substitution reduction still cannot be reduced by loop detection.

\end{document}